\documentclass[12pt]{article}

\usepackage[latin1]{inputenc}

\usepackage{latexsym}

\usepackage{overcite}

\makeatletter \def\@cite#1{\textsuperscript{(#1)}} \makeatother

\newcommand{\sg}{\sigma}
\newcommand{\defi}{\stackrel{Df}{=}}
\newcommand{\den}{\stackrel{d}{=}}
\newcommand{\rep}{\stackrel{\wedge}{=}}

\newenvironment{proof}{  {\bf Proof:}}

\newenvironment{remark}{ {\bf Remark:} \begin{quotation}} {\end
 {quotation}}

\newenvironment{remarks}{ {\bf Remarks:} \begin{quotation}}{\end
 {quotation}}

\newenvironment{des}{\begin{enumerate} \begin{quotation}}
{\end{quotation} \end{enumerate}}

\newtheorem{ax}{A.}
\newtheorem{df}{Def.}
\newtheorem{tre}{Thm.}

\title{AXIOMATIC FOUNDATIONS OF GALILEAN QUANTUM FIELD THEORIES}

\author {G. Puccini\\Dep. de F\'\i{}sica, Universidad
Nacional de La Plata, Argentina\thanks{Present address: Instituto de
 Neurociencias, Universidad Miguel Hern\'andez, Apartado
18, 03550 San Juan de Alicante, Spain.},\\
H. Vucetich\\ Instituto de F\'\i sica\\
Universidad Nacional Aut\'onoma de M\'exico\\
A. Postal 70-543, M\'exico D.F. 04510, M\'exico\thanks{On leave of
  absence from Observatorio Astron\'omico, Universidad
Nacional de La Plata, Paseo del Bosque S./N., (1900) La Plata,
Argentina}
}

\begin{document}
\maketitle

\begin{abstract}
A realistic axiomatic formulation of Galilean Quantum Field Theories
is presented, from which the most important theorems of the theory can be
deduced. In comparison with others formulations, the formal aspect has
been improved by the use of certain mathematical theories, such as group
theory and the theory of rigged Hilbert spaces. Our approach
regards the fields as real things with symmetry properties. The general
structure is analyzed and contrasted with relativistic theories.
\end{abstract}

KEYWORDS: Axiomatization, Quantum Field Theories, Galilean Invariance.

\section{INTRODUCTION}

There are different ways to present a physical theory. On the one
hand, the {\em informal} way (for example, the historical approach)
proceeds with the help of analogies and metaphors. These approaches
are hardly appropriate for the study of the structure of the
theory. In fact, in these formulations the structure emerges
gradually, the assumptions are rarely exposed explicitly and the basic
concepts are presented disorderly. Moreover, the theory is interpreted
by analogies and heuristic clues. On the other hand, the {\em formal}
way, or the axiomatic approach as it is most widely known, consists of
a logical organization of the theory with both an adequate
characterization of the physical meaning of the symbolism and an
examination of the metatheoretical aspects. In our opinion, every
formal analysis of the structure of a theory should be a consequence
of the study of its axiomatic framework in such a way that all the
presuppositions, theorems and interpretation rules are exhibited
explicitly.

There are different informal approaches for the quantization of a
nonrelativistic field theory. Redmond and Uretzky \cite{RU62}
developed the in-out formalism in the case of second
quantization. Dresden and Kahn \cite{DK62} studied non relativistic
field theories assuming invariance with respect to the Euclidean group
of transformations. Schweber \cite{SS62} studied the nonrelativistic
Bethe-Salpeter equation. L\'evy-Leblond \cite{JMLL67} investigated
field theories based on the notion of field operators, local
commutability and Galilean invariance, and Dadashev \cite{Dada85}
presented a system of axioms following the axiomatics of Wightman for
the relativistic case.

In this article we present an axiomatic formulation of the Galilean
quantum field theories (GQFT) and we analyze its conceptual and
mathematical structure. In our axiomatization we follow the basic
postulates adopted by many of the formulations of the relativistic
theory but we adapt them to the Galilean framework, i.e. Galilean
covariance, local commutability and irreducibility. The mathematical
aspect has been improved by the use of certain mathematical theories,
such as group theory and the theory of rigged Hilbert
spaces. Differing from the bulk of the formulations, we assume a
realistic ontology, namely that of references \cite{BungeT3,BungeT4}
and we construct the theory from the notion of {\em basic thing} and
{\em system with properties}. In the present case, these are: {\em
basic fields} and {\em systems of basic fields} with different
properties (symmetry properties among them), respectively. Besides, we
base our interpretation of the theory in the rigorous semantics of
references \cite{BungeT1,BungeT2}.

From these postulates we will obtain rigorous and general proofs that
show the rich structure of the theory. For example, we will see that
Galilean quantum field theory is less restrictive than its
relativistic counterpart, main\-ly due to the restricted form of the
commutation relations of the local field operators. As a consequence
of this weaker condition, the powerful results of relativistic field
theory, such as CPT and spin-statistics theorems, cannot be derived in
GQFT. This shows the important role that commutation relations play in
relativistic field theories. However, we must remark that there are
several powerful results in Galilean field theory that do not exist in
the relativistic one, such as the existence of a mass superselection
rule.

The structure of the paper is the following. In the second section, we
expound some ontological concepts that must be presupposed by any
quantum field theory. In the third section we review the Galilei
group, its Lie algebra and its physical representations. The axiomatic
of GQFT is develop in the fourth section, along with remarks that
clarify the axioms and some of the principal theorems. In the fifth
section we discuss some of the consequences of GQFT such as crossing
symmetry and spin-statistic theorems, and the final conclusions are
given in section sixth.

\section{ONTOLOGICAL BACKGROUND}

It is undeniable that there are some ontological queries that must be
carefully elucidated in the background of any field theory. In order
to understand the physical concepts of ``particle", ``field" and their
mutual actions it is mandatory to have an accurate characterization of
the concepts of {\em system, interaction} and {\em state}. Since we
deal with physical systems, we are interested in a formal
characterization of the ontological concept of a system first.  We
shall assume the realistic ontology of Bunge\cite{BungeT3,BungeT4} and
in this section we shall summarize the results relevant for this
paper. 

The basic concept of this ontology i.e. that of a {\em substantial
individual}, will be denoted by $x$. Substantial individuals can
associate to form new substantial individuals, and they differ from
the fictional entities called {\em bare individuals} (i.e. individuals
without properties) precisely in that they have a number of properties
$P$ in addition to their capability of association. We shall propose
some useful definitions.

\begin{df}[Substantial Property]
The substantial properties are those possessed by some real thing, i.e.
$$
P\in {\cal P} \leftrightarrow(\exists x)(x\in S\wedge Px) .
$$
\end{df}
Here ${\cal P}$ is the set of all substantial properties, $S$ is
the set of all substantial individuals and $Px$ designate `$x$ has the property
$P$'.

\begin{df}
The set of all the properties of a given individual is given by
$$
P(x) = \{P\in {\cal P} | Px\}.
$$
\end{df}

These properties can be {\em intrinsic} ($P_i$) or {\em relational}
($P_r$). The intrinsic properties (e.g. charge, spin) are inherent and
they are represented by unary predicates or  applications while
relational properties (e.g. position, momentum) are represented by
$n$-ary predicates ($n > 1$), as long as nonconceptual arguments are
considered.

We will need below the concepts of same and identical individuals:
\begin{df}
Two individual are the same if they have exactly the same properties,
i.e.,
$$
\forall x,y \in S, P(x)=P(y) \Rightarrow x \equiv y .
$$
\end{df}

\begin{df}
Two individual are identical if their intrinsic properties are the
same, i.e.,
$$
\forall (x,y \in S) [P_i(x)=P_i(y) \Rightarrow x
\stackrel{id}{\leftrightarrow}y] .
$$
\end{df}
For example, two spin-up electrons are identical because they have
the same intrinsic properties. We now  can define a concrete thing as
made up from substantial individuals $x$ together with their
properties $P(x)$.

\begin{df}[Thing]
Denoting a thing by $X$, we define it as the ordered pair
\footnote{\rm The brackets $\langle, \rangle$ denote an ordered set.}:
$$
X=\langle x, P(x) \rangle .
$$
\end{df}

The set of all things will be denoted by $\Theta $.  Some things are
complex, so we can introduce the notion of composition of things:
\begin{df}[Absolute Composition]
For every $X \in \Theta $, the composition of $X$ is:
$$
{\cal C}(X)= \{Y \in \Theta | Y \sqsubset X \}
$$
where ``$Y \sqsubset X $'' designates ``$Y$ is a part of $X$''.
\end{df}

Let us introduce the concept of action:
\begin{df}[Action]
A thing $X$ acts on another thing $Y$ if $X$ modifies the behavior of
$Y$.  ($X \rhd Y : X $ acts on $Y$).  If the action is mutual it is
said that they interact ($X \Join Y $).
\end{df}

\begin{df}[Connection]
Two things are connected  if at least one of them acts on the other.
\end{df}

Now we have all we need to define a concrete system:
\begin{df}[System]
A {\em system} $\sg$ is a thing composed of at least two different
connected things.
\end{df}

Two important features of a system are its composition and its environment,
\begin{df}[A-composition] The composition of a system $\sg$ at a given
time $t \in T$ with respect to a class $A$ of things is the set of its
$A$-parts at $t$:
$$
{\cal C}_A(\sg,t)= \{X \in A | X \sqsubset \sg \mbox{at} t \in T\}
$$
\end{df}

\begin{df}[A-environment]
The environment of a system $\sg$ at time $t$ is the set of things of
kind $A$ that are not components of $\sg$ but that are connected with
some or all the components of $\sg$, that is,
$$
{\cal E}_A(\sg,t)=\{ X \in A | X \not \in {\cal C}_A(\sg,t) \wedge
(\exists Y)(Y \in {\cal C}_A(\sg,t) \wedge (X \rhd Y \vee Y \rhd X))\}
$$
\end{df}

\begin{df}[Closed System]
A system is closed at the instant $t$ if and only if its environment
is empty, i.e. ${\cal E}_A(\sg,t)= \emptyset$
\end{df}

Theoretical physics does not characterize concrete things but rather
concepts, in particular with conceptual schemes called models of
things.  We will assume that any property $P$ of a thing $X$ it is
represented by a mathematical application $F$, that is to say,
$F\stackrel{\wedge} {=} P $.

\begin{df}[Functional schema]
Let $\sg$ be a system. A functional schema $b$ of $\sg$ is a certain
nonempty set $M$, together with a finite sequence of mathematical
applications $F_i $ on $M$, each one of which represents a property of
$\sg$.  Shortly:
$$
b=\langle M,{\bf F} \rangle \;\mbox{where}\;{\bf F}=\langle
F_1,F_2,\ldots, F_p \rangle:
M \rightarrow V_1 \times V_2 \times \cdots \times V_p
$$
\end{df}

Therefore, the system $\sg$ will be represented by the functional
schema $b$, that is to say, $b \stackrel {\wedge} {=} \sg$.

It is natural to assume that all things are in some state. The state
of a system can be characterized as follows:
\begin{df}[State function]
\label{estado}
Let $\sg$ be a system modeled by a functional sche\-ma $b = \langle
M,{\bf F} \rangle $. Then, each $F_i $ is a {\em state
function} of $\sg$.  ${\bf F}$ is the {\em state vector} of $\sg$, and
its value
$$
{\bf F} (m)=\langle F_1,F_2,\ldots, F_p \rangle(m)=\langle
F_1(m),F_2(m),\ldots, F_p (m)\rangle
$$
for any $m \in M$  {\em represents the state} of $\sg$ in the
representation $b$.
\end{df}

In our axiomatic we shall be interested with basic fields and systems of
basic field. These will be the specific physical systems to be
characterized by our axiomatic frame.

\section{MATHEMATICAL BACKGROUND}

The symmetry group of nonrelativistic transformations from one
inertial frame of reference to another is the proper Galilei group
$G$, defined as
\begin{df}
The set of all transformations in Galilean space-time consisting
of space and time displacement, pure Galilei transformations and space
rotations constitutes a continuous group $G$ of $10$ parameters called
the {\em Galilei group}.

The generic element will be denoted by $g=(b,{\bf a},{\bf v},R) $,
with $b\den$ time translation, ${\bf a} \den$ space translation, ${\bf
v} \den$ pure Galilean transformation and $R\den$ three-dimensional
rotation. Its action on space and time is represented by the
coordinate transformation
$$
{\bf x} \rightarrow {\bf x}'=R{\bf x}+{\bf v}t+{\bf a},
$$
$$
t \rightarrow t'=t+b.
$$
The multiplication law is given by
$$
g'g=(b', {\bf a}',{\bf v}',R')(b,{\bf a}, {\bf v} ,R)
$$
$$
=(b'+b, {\bf a}'+R'{\bf a}+b{\bf v}', {\bf v}'+ R'{\bf v}, R'R).
$$
The identity element is
$$
e=(0,{\bf 0},{\bf 0},1).
$$
The inverse element is
$$
g^{-1}=(-b,-R^{-1}({\bf a}-b{\bf v}),-R^{-1}{\bf v},R^{-1}).
$$
\end{df}

Contrary to the case of the Poincar\'e group, In\"on\"u and Wigner
\cite{InWig52} found that the basis vectors of the ordinary
representation of the Galilei group cannot be interpreted as physical
states of particles. On the other hand, Bargmann \cite{Barg54} had
shown that the physically meaningful representations need\-ed to
describe particles are the projective representations. In other words,
we obtain a unitary representation of the symmetry group $G$ if to
each element $g \in G$ (indeed of its universal covering group) we can
associate a unitary operator $\hat U(g)$ on the physical Hilbert
space. However, the composition rule of the unitary operators $\hat
U(g)$ and $\hat U(g')$ cannot be written as in the case of an {\em
ordinary representation}, $\hat U(g) \hat U(g') = \hat U(g g').$
Indeed, the group is represented {\em projectively} on physical
states; that is, the unitary operators satisfy the composition rule,
$\hat U(g) \hat U(g') = e^{i\zeta( g,g')} \hat U(g g')$, with $\zeta$
a real phase. Bargmann \cite{Barg54} has shown that for the Galilei
group the phases $\zeta( g,g')$ cannot, in general, be made equal to
zero by a redefinition of $\hat U(g)$.

The presence of phases in a projective representation of a group has a
counterpart in the Lie algebra of the group: it is the appearance of
terms on the right-hand side of the commutation relations proportional
to the unit element. These terms are called {\em central charges}. If
the generators $\hat H,\hat P_i,\hat K_i,\hat J_i$ denote time
translations, space translations, pure Galilean transformations and
rotations respectively, we can define:
\begin{df}
\label{liealg}
The Lie algebra ${\cal G}$ of the Galilei group is given by the
commutation rules of the generators $\hat H,\hat P_i,\hat K_i,\hat
J_i$ which satisfy the following relations:
$$
[\hat J_i,\hat J_j]=i \hbar \epsilon_{ijk} \hat J_k,[\hat
J_i,\hat K_j]=i \hbar \epsilon_{ijk} \hat K_k, [\hat J_i,\hat
P_j]=i \hbar \epsilon_{ijk} \hat P_k,
$$
$$
[\hat K_i,\hat H]=i\hbar \hat P_i,[\hat K_i,\hat P_j]=i \hbar
\delta_{ij}m,
$$
$$
[\hat J_i,\hat H]=0,[\hat K_i,\hat K_j]=0,[\hat
P_i,\hat P_j]=0,[\hat P_i,\hat H]=0,
$$
\end{df}
where $\hbar$ is Planck's constant and $m$ is a real number (the
central charge).

So, the Galilean algebra does allow for a central charge that cannot
be removed by a redefinition of the generators \cite{Barg54}. However,
we can enlarge the Galilean group $G$, by adding one more generator to
its Lie algebra, which commutes with all the other generators, and
whose eigenvalues coincide with the central charge. This generator
will be denoted by $\hat M$. This extension by an Abelian
one-dimensional group is called {\em central} because $\hat M$
commutes with all the other elements of the Lie algebra, and it is
nontrivial because $\hat M$ appears on the right side of some
commutations relations. The expanded Lie algebra $\tilde {\cal G}$ is
free of central charges and the extended group ${\tilde G}$ has only
ordinary representations. The space and time translations form an
abelian subgroup of the Galilei group. The representations of this
subgroup are known and will be denoted by their eigenvalues (${\bf
p}$, $E$).

The following theorems can be deduced from the above definitions:
\begin{tre} The algebra $\tilde{\cal G}$ admits the following invariants:
\begin{eqnarray*}
\hat Q_1 &=&\hat M,\\
\hat Q_2 &=&2 \hat M \hat H - \hat {\bf P}^2 \equiv 2 \hat M \hat
W,\\
\hat Q_3 &=&(\hat M \hat {\bf J} - \hat {\bf K} \times \hat {\bf P})^2
= \hat M^2 \hat {\bf S}^2,
\end{eqnarray*}
where $\hat W$ is the {\em internal energy operator} and $\hat{\bf S}$
is the {\em spin operator}.
\end{tre}

\begin{proof}
See references \cite{Hamm60,JMLL63}.
\end{proof}

\begin{tre}
\label{repre}
\begin{description}
\item a) The set of basis vectors $ \{ | {\bf p} ,E,\lambda \rangle \}
$ spans a vector space which is invariant under Galilei group
transformations. The action of an arbitrary Galilei transformation on
these basis vectors is given by:
$$
\hat U(g) |{\bf p},E,\lambda \rangle = \exp [ -\frac{i}{\hbar}(E' b
-{\bf p}'.{\bf a}) ] \sum_{\lambda'} | {\bf p}',E',\lambda' \rangle
D_{\lambda' \lambda}^{(s)}(R) ,
$$
\begin{eqnarray*}
{\bf p}' &=& R{\bf p}+ m {\bf v},\\
E' &=& { \frac{{\bf p'}^2}{2m} } + W,\\
&=& E + {\bf v}.R{\bf p}+{\frac{1}{2}}m {\bf v}^2,
\end{eqnarray*}
and $D^{(s)}(R)$ is the representation matrix of the rotation group
corresponding to the spin $s$.
\item b) The resulting representation, labeled by $[m,W,s]$, is
unitary and irreducible.
\end{description}
\end{tre}

\begin{proof}
See references \cite{Barg54,JMLL63}.
\end{proof}

\begin{tre}
\label{parab}
\mbox{} Every irreducible representati\-on of the translation subgroup
$({\bf p},E)$ satisfy the following condition:
$$
E - \frac{{\bf p}^2}{2m} = W=const.
$$
\end{tre}

\begin{proof}
See references \cite{Wigh62,JMLL63}
\end{proof}

\begin{tre}
The invariant measure is given by,
$$d\mu({\bf p},E)=  \delta(E - \frac{{\bf p}^2}{2m}- W)d^3p  dE$$
\end{tre}

\begin{proof}
From {\bf Thm.\ref{parab}}.
\end{proof}

\mbox{}

We shall use below the Peter-Weyl theorem, which is a generalization
of the Fourier theorem and holds for every compact Lie group (see
\cite{TungGT}). Although the Galilean group is non-compact,
it is valid for the compact part, i.e. the representation of the
rotation group:
\begin{tre}
The irreducible representation basis functions form a complete basis
in the space of (Lebesgue) square-integrable functions defined on the
group manifold.
\end{tre}

In our axiomatic formulation, the generators of the Lie algebra
$\tilde {\cal G}$ will be postulated and identified later on by means
of semantic assumptions.

\section{AXIOMATICS OF GQFT}

In this section we shall exhibit the axiomatic structure of the theory
based on the ontological background and making use of the symmetry
properties presented above. Firstly, we shall present the two set of ideas
that GQFT takes for granted: Formal and material background. The formal
background consist of all the logical and mathematical ideas it employs. The
material background consists of all the generic and specific theories it
presupposes. \footnote{We shall display only a list of the
main items.}

\subsection{Formal Background}
\begin{enumerate}
\item Classical logic.
\item Formal semantics \cite{BungeT1,BungeT2}.
\item Mathematical analysis with its presuppositions, and the theory
of generalized functions \cite{GelfandGF}.
\item Group theory.
\end{enumerate}

\subsection{Material Background}

\begin{enumerate}
\item Protophysics \cite{BungeFP} (i.e. Physical probabilities, Chronology,
Physical geometry,\ldots).
\item Dimensional analysis.
\end{enumerate}

\begin{remark}
  We do not use dimensional analysis explicitly in this paper (except
in axiom \ref{PhysProp}). However, dimensional analysis does play a
very important, albeit silent, role in every physical theory: it
classifies physical quantities in dimensionally homogeneous classes
and so restricts the form of their mathematical representations.
\end{remark}

\subsection{Primitive Basis}
The conceptual space of the theory is generated by the basis {\bf B}
of primitive concepts, where
$$
{\bf B}={\langle E_{3},T, \overline{\Sigma}, \Sigma, {\cal H}_E, {\cal
P},A, \hbar , {\cal F}, G\rangle}
$$

The elements of this basis will be characterized mathematically
{\bf[M]}, physically {\bf [P]} and semantically {\bf [S]} by the
axiomatic basis of the theory and the derived theorems.

\subsection{Definitions}
\begin{df}
$K \defi $ set of physical reference systems.
\end{df}

\begin{df}
eiv $\hat A \defi$ eigenvalues of $\hat A$.
\end{df}

\begin{df}
$ \langle \Phi_1 | \Phi_2 \rangle \defi $ scalar product of the
vectors $\Phi_1$ and $\Phi_2$.
\end{df}

\subsection{Axioms}

\subsubsection*{Group I: Space and Time}

\begin{ax}[Space]
\label{espa}
\begin{des}
\item {\bf [M]} $E^{3} \equiv $ Euclidean three-dimensional space.
\item {\bf [S]} $E^{3} \rep $ physical space .
\end{des}
\end{ax}

\begin{ax}[Time]
\label{tiem}
\begin{des}
\item {\bf [M]} $T \equiv $ interval of real line $\Re$.
\item {\bf [S]} $T \rep $ time interval.
\item {\bf [S]} The relation $\leq$ that orders $T$ means ``before"
$\vee$ ``simultaneous with".
\end{des}
\end{ax}

\begin{remark}
  These axioms only characterize our \emph{use} of the notions of
  space and time. Ontological analysis of these notions can be seen in
  references \cite{BungeT3,etb}
\end{remark}

\subsubsection*{Group II: F-Systems and States}

\begin{ax}[Systems]
\label{sist}
\begin{des}
\item {\bf [M]} $\Sigma, \overline \Sigma $: nonempty numerable sets.
\item {\bf [S]} $(\forall{\sg_i})_{\Sigma} (\sg_i \den $ {a basic
field}).
\item {\bf [S]} $(\forall{\sg})_{\Sigma} ({\cal
C}(\sg)=\{\sg_1,\ldots{},\sg_N \}) (\sg \den \mbox{f-system}).$
\item {\bf [S]} $(\forall{\overline \sg})_{\overline \Sigma}
(\overline \sg \den \mbox{environment of some f-system })$. {In
particular,} $ (\overline \sg_o \den$ {the empty environment}).
\item {\bf [S]} $(\forall{\sg})_{\Sigma} (\langle \sg, \overline \sg_o
\rangle \den \mbox{a closed f-system}).$
\item {\bf [P]} $(\exists K) (K \subset \overline \Sigma \wedge$
the configuration of $ k \in K$ {is independent of time}).
\end{des}
\end{ax}

\begin{ax}[State Space]
\label{espest}
\begin{des}
\item {\bf [M]} $(\forall \sg)_{\Sigma} (\exists {\cal H}_E=\langle
{\cal L,H,L'}\rangle \equiv \mbox{rigged Hilbert space}). $
\item {\bf [P]} \mbox{} There exists a one-to-one correspondence between
physical states of $\sg \in \Sigma $ and rays ${\cal R}_{\sg} \subset
{\cal H}$.
\item {\bf [M]} $(\forall \sg)_{\Sigma} ({\cal
C}(\sg)=\{\sg_1,\ldots{},\sg_N \} \Rightarrow {\cal H}_E =
\otimes^{N}_{i=1} {\cal H}_{E_i} ).$
\item {\bf [S]} $|\Phi(\sg,k)\rangle \in {\cal R}_{\sg} \den $ state
vector that is the representative of the ray ${\cal R}_{\sg}$ that
corresponds to the f-system $\sg$ with respect to $k \in
K$.
\footnote{\rm To avoid unnecessary complexity in notation we
are not going to make explicit the dependence of the state on the
system and on the reference system.}
\item {\bf [P]} $(\exists |0 \rangle)_{{\cal R}_{\sg}} ( |0 \rangle
\den $ the normalized state called the vacuum state $)$.
\end{des}
\end{ax}

\begin{remarks}
\begin{enumerate}
\item The third axiom must be explicitly postulated in order to state
that the reference class of the GQFT is not an empty set. Moreover,
the elements of this set are semantically interpreted as
`fields'. These axioms are not trivial since that a different
interpretation could claim that GQFT deals olny with experimental
results (i.e., a set of data not a physical system) or that the
reference class are particles (not fields) because they are
observables.
\item The rigged Hilbert space is an extension of the ordinary Hilbert
space and it is introduced here in order to include the operators (as
position and linear momentum operators) whose eigenfunctions does not
have finite norm. Moreover, from the physical point of view, the axiom
states that every physical state of the system is mapped on a subset
$\cal H$.
\end{enumerate}
\end{remarks}

\subsubsection*{Group III: Operators and Physical Quantities}

\begin{ax}[Physical Properties] \label{PhysProp}
\begin{des}
\item {\bf [M]} ${\cal P}\equiv \mbox{nonempty family of applications
over } \Sigma$.
\item {\bf [M]} $A \equiv \mbox{ ring of operators over } {\cal H}_E$.
\item {\bf [P]} $(\forall P)_{\cal P} (\exists \sg)_\Sigma (P \in
P(\sg))$.
\item {\bf [P]} $(\forall P)_{\cal P} (\exists \hat A)_A(\hat A \rep
P).$
\item {\bf [P]} ${(\forall \sg)}_{\Sigma} {(\forall \hat A)}_A
{(\forall a)}_{\Re} (eiv \hat A=a \wedge \hat A \rep P \Rightarrow a$
is the sole value that $P$ takes on $\sg )$.
\item {\bf [M]} $\hbar \in \Re^{+}.$
\item {\bf [P]} $[\hbar]=LMT^{-1}.$
\end{des}
\end{ax}

\begin{ax}[Linearity and Hermiticity]
\begin{des}
\item {\bf [M]} ${(\forall \sg)}_{\Sigma} (\forall \hat A)_A (\forall
P)_{\cal P} (\hat A \rep P \wedge |\Phi_1 \rangle, |\Phi _2 \rangle
\in {\cal H}_E \Rightarrow $

\begin{enumerate}
\item $\hat A [\lambda_1 |\Phi_1 \rangle + \lambda_2 |\Phi_2 \rangle
]= \lambda_1 \hat A |\Phi _1 \rangle + \lambda_2 \hat A |\Phi_2
\rangle, \mbox{ with } \lambda_1, \lambda_2 \in {\cal C}$
\item $\hat A^{\dag}=\hat A$
\end{enumerate}
\end{des}
\end{ax}

\begin{ax}[Probability Densities]
\begin{des}
\item {\bf [P]} $(\forall \sg)_{\Sigma} (\forall \hat A)_A (\forall
P)_{\cal P} (\forall |a \rangle)_{{\cal H}_E} (\forall |\Phi \rangle
)_{{\cal H}_E} (\hat A \rep P \wedge \hat A |a \rangle=a |a \rangle
\Rightarrow $ the probability density $\langle \Phi |a \rangle \langle
a | \Phi \rangle $ corresponds to the property $ P $ of the f-system $
\sg )$.
\end{des}
\end{ax}

\begin{ax}[Unitary Operators]
\begin{des}
\item {\bf [P]} $(\forall \sg)_{\Sigma } (\forall \hat A)_A (\forall
P)_{\cal P} (\forall \hat U)(\hat A \rep P \wedge \hat U $ is an
operator on $ {\cal H}_E \wedge \hat U^{\dag} =\hat U^{-1} \Rightarrow
\hat U \hat A \hat U^{-1} \rep P)$.
\end{des}
\end{ax}

\begin{remark}
The axioms of Group III are typical of a realistic formulation of
Quantum Mechanics.  In addition of the mathematical properties
(attributes) of the operators, the axioms characterize semantically
the properties of the systems, that is, every property of the system
is represented by an operator. Note that the eigenvalues of these
operators are not interpreted as the only possible values of a
`measurement' made on the system, since the relationship with
measurement cannot be made at the level of principles.
\end{remark}

\subsubsection*{Group IV: Quantum Fields}

\begin{ax} [Field Operators]
\label{field}
\begin{des}
\item {\bf [M]} ${\cal F}\subset A \equiv $ nonempty set of operators
over ${\cal H}_E$.
\item {\bf [S]} $(\forall \sg_i)_{\Sigma} (\exists \hat \psi_\lambda
)_{\cal F} (\hat \psi_\lambda \den \lambda$-component of the local
field operators associated with the basic field $\sg_i )$.
\item {\bf [S]} $(\forall \sg_i)_{\Sigma} (\forall ({\bf x},t) )_{E_3
\times T} (\hat \psi_\lambda ({\bf x},t) \rep $ the amplitude of the
basic field $\sg_i $ at ${\bf x},t)$ .
\item {\bf [P]} $(\forall \hat \psi_\lambda)_{\cal F} (\forall ({\bf
x},t) )_{E_3 \times T} (\hat \psi_\lambda ({\bf x},t) |0 \rangle =0
)$.
\item {\bf [P]} $(\forall \hat \psi_\lambda)_{\cal F} (\forall ({\bf
x},t) )_{E_3 \times T} (\langle 0 | \hat \psi_\lambda ({\bf x},t) =
\langle x,t; \lambda | )$.
\end{des}
\end{ax}

\begin{ax}[Irreducibility]
\label{oper}
\begin{des}
\item {\bf [M]} The set of field operators is irreducible, i.e.:
\begin{eqnarray*}
&(\forall \sg)_{\Sigma}(\forall \hat O)_A\left\{
\hat O =  \sum_{N,M =0}^{\infty} \right.& \\
&\left.\sum_{\lambda'_1\ldots{}\lambda'_N}
\sum_{\lambda_1\ldots{}\lambda_M}\int {\cal D}{\bf x}'{\cal D}{\bf x}
{\hat
\Psi}_{\lambda'}^{\dag}({\bf x}',t){\hat \Psi}_{\lambda}({\bf x},t)
{\cal C}_{NM}\right\}&
\end{eqnarray*}
where
$$
{\cal D}{\bf x}'{\cal D}{\bf x} \den d^3 x'_1\ldots{}d^3 x'_N d^3
x_1\ldots{}d^3 x_M ,
$$
$$
{\hat \Psi}_{\lambda'}^{\dag}({\bf x}',t) \den {\hat
\psi}_{\lambda'_1}^{\dag}({\bf x}'_1,t)\ldots{}{ \hat
\psi}_{\lambda'_N}^{\dag}({\bf x}'_N,t),
$$
$$
{\hat \Psi}_{\lambda}({\bf x},t) \den \hat \psi_{\lambda_1}({\bf
x}_1,t)\ldots{} \hat \psi_{\lambda_M}({\bf x}_M,t),
$$
$$
{\cal C}_{NM} \den C_{\lambda'_1\ldots{}\lambda'_N
\lambda_1\ldots{}\lambda_M}({\bf x}'_1,\ldots{}, {\bf x}'_N, {\bf
x}_1,\ldots{},
{\bf
x}_M),
$$
$$
{\cal C}_{NM}={{\cal C}_{MN}}^*,
$$
and $\hat \psi_{\lambda_i} \den \lambda$-component of the $i$-th field
operator $\hat \psi$.
\end{des}
\end{ax}

\begin{ax}[Local Commutativity]
\label{causa}
\begin{des}
\item {\bf [M]} The field operators satisfy the following commutation or
anticommutation rules at equal times:
$$
[\hat \psi_\lambda({\bf x},t) ,\hat \psi_{\lambda'}({\bf
y},t)]_{\mp}=[{\hat \psi^{\dag}}_\lambda({\bf x},t) ,{\hat
\psi^{\dag}}_{\lambda'}({\bf y},t)]_{\mp}=0
$$
$$
[\hat \psi_\lambda({\bf x},t) ,{\hat \psi^{\dag}}_{\lambda'}({\bf
y},t)]_{\mp}=\delta_{\lambda \lambda'} \delta^3({\bf x}-{\bf y}).
$$
\end{des}
\end{ax}

\begin{remarks}
\begin{enumerate}
\item $\Sigma \times \overline \Sigma$ is the class of reference of
GQFT. A member $\langle \sigma, \overline \sigma \rangle$ denotes an
arbitrary system (so called f-system) composed of basic fields
$\sigma_i$, eventually in interaction with the environment $\overline
\sigma$.  In particular, we will deal with closed f-systems which are
described by $\langle \sigma, \overline \sigma_o \rangle$, with
$\sigma_o$ denoting the empty environment. These physical fields are
characterized mathematically by operators acting over a region of
space and time (see {\bf A.\ref{field}}) and with certain
transformation properties (see {\bf A.\ref{ftran}}). In other words,
{\bf A.\ref{sist}} characterizes the physical system that the theory
refers to, and {\bf A.\ref{field}} characterizes the mathematical
concept of a field.  Indeed, not all mathematical field must represent
a physical (real) entity.  However, we will assume that the field
operator represent a physical property of the basic field, that is,
the amplitude of $\sigma_i$.
\item We must emphasize that the {\em primitive concept} of this
field theory is the {\em physical field} that will be characterized
(its properties) by an operator. As we shall see, the concept of
``particle" is a {\em derived concept} as a the ``quantum" of the
field.
\item Note that {\bf A.\ref{field}} states that ${\cal F} \subset
A$. Indeed, $\cal F$ is a particular set that need to be considered
independently of $A$. Moreover, {\bf A.\ref{oper}} states that the
properties of a system of fields will be represented as a polynomial
of the field operators associated to each basic field.
\item A basic field $\sigma_i$ is modeled by a functional schema $b=\langle M,
{\bf F} \rangle$, where we can identify the set $M$ as the Cartesian
product of certain set $\Sigma \times \overline \Sigma \times E_3 \times T
\times A$ and ${\bf F}$ as the set of
operators $\langle \hat \psi_{-s} ({\bf x},t),\ldots,\hat \psi_{s} ({\bf
x},t)\rangle$. Moreover,
the physical space of accessible states is $V_i={\cal H}_S \oplus {\cal H}_A
\subset {\cal H}_E$ (see {\bf A.\ref{espest}} and {\bf Thm.\ref{acce}}).
\item Since the reference of nonrelativistic quantum field theories
are basic fields, in order to obtain a unitary quantum theory of
fields, we need only one equation of motion. This situation is
different from that of the classical theories of fields, where one
equation of motion for fields and another for particles are needed.
\end{enumerate}
\end{remarks}

\subsection{Definitions}
\begin{df} [Vector basis of an f-system $\sg$] \mbox{}
\begin{displaymath}
|q_1,q_2,\ldots{},q_N \rangle = |q_1 \rangle |q_2 \rangle\ldots{}|q_N
 \rangle,
\end{displaymath}
where $|q_i \rangle \den |{\bf x}_i,t;\lambda_i \rangle.$
\end{df}

\begin{df}[Symmetric Hilbert spaces]
\begin{eqnarray*}
&{\cal H}_S \equiv \{ |q_1, q_2,\ldots{}, q_N \rangle \mid |q_1
,q_2,\ldots{}, q_N \rangle \in {\cal H}_E&\\
& \wedge\; |q_1,q_2,.., q_N
\rangle = |q_{{\cal P}1},q_{{\cal P}2},\ldots{},q_{{\cal P}N}\rangle
\}.&
\end{eqnarray*}
\end{df}

\begin{df}[Antisymmetric Hilbert spaces]
\begin{eqnarray*}
&{\cal H}_A \equiv \{|q_1, q_2,\ldots{}, q_N \rangle \mid |q_1,
q_2,\ldots{},
q_N \rangle \in {\cal H}_E&\\
& \wedge\; |q_1,q_2,.., q_N \rangle = (-1)^{\mu}
|q_{{\cal P}1},q_{{\cal P}2},\ldots{},q_{{\cal P}N}\rangle \}&
\end{eqnarray*}
 with
${\mu}$ the number of permutations ${\cal P}$ of $|q_i \rangle$.
\end{df}

\subsection{Theorems}
\begin{tre}
\label{gral}
The general basis vector of an f-system $\sg$ can be written in terms
of operators ${\hat \psi^{\dag}}$ acting on the vacuum state as
follows:
\begin{displaymath}
|q_1,q_2,\ldots{},q_N \rangle = \hat \psi^{\dag}(q_1)\hat
\psi^{\dag}(q_2)\ldots{} \hat
 \psi^{\dag}(q_N) |0 \rangle,
\end{displaymath}
 where ${\hat \psi^{\dag}}(q_i) \den {\hat
\psi^{\dag}}_{\lambda_i}({\bf
x}_i,t)$.
\end{tre}

\begin{proof}
From {\bf A. \ref{espest}} and {\bf A. \ref{field}}.
\end{proof}

\begin{tre}
\label{sime}
The general basis vector $|q_1,q_2,\ldots{},q_N \rangle$ of an f-system
$\sg$ can be written as:
$|q_1,q_2,\ldots{},q_N \rangle= (\pm)^{\mu} |q_{{\cal P}1},q_{{\cal P}2},
\ldots{},q_{{\cal P}N}\rangle\;,$ where ${\mu}$ is the number
of permutations of the operators $\hat \psi^{\dag}$ associated to {\em
identical} basic fields $\sg_i$. 
\end{tre}

\begin{proof}
From {\bf Thm. \ref{gral}} and {\bf A. \ref{causa}}.
\end{proof}

\begin{tre}
\label{acpsi}
The field operator $\hat \psi_{\lambda}({\bf x},t) $ acts on the basis
vector $| {\bf x},t;\lambda \rangle$ as:
$$
\hat \psi_{\lambda}({\bf x},t) | {\bf x},t;\lambda \rangle =
| 0 \rangle.
$$
\end{tre}

\begin{proof}
Using {\bf A. \ref{espest}}, {\bf A. \ref{field}} and {\bf
A. \ref{causa}}.
\end{proof}

\begin{tre}
The action of operator $\hat \psi (q) \den \hat \psi_{\lambda}({\bf
x},t)$ on a general basis vector $|q_1,q_2,\ldots{},q_N \rangle$ is given
by,
\begin{displaymath}
\hat \psi (q) |q_1, q_2,\ldots{}, q_N \rangle = \sum_{r=1}^N (\pm)^{r+1}
\delta(q-q_r) |q_1, \ldots{} ,q_{r-1},q_{r+1}, \ldots{} ,q_N \rangle
\end{displaymath}
with $+1$ or $-1$ for $ |q_1, q_2,\ldots{}, q_N \rangle \in {\cal H}_S$
and ${\cal H}_A$ respectively.
\end{tre}

\begin{proof}
From {\bf Thm. \ref{acpsi}} or {\bf A. \ref{causa}}.
\end{proof}

\begin{tre}[Orthonormality]
\label{norm}
The following orthonormality condition holds for a general basis vector:
$$\langle q'_1, q'_2,\ldots{}, q'_N |q_1, q_2,\ldots{}, q_M
\rangle = \delta_{NM} \sum_{\mu} (\pm 1)^{\cal P} \prod_i \delta (q_i -
q'_{{\cal P}_{i}})
$$
\end{tre}

\begin{proof}
From {\bf Thm. \ref{gral}}, {\bf Thm. \ref{sime}} and {\bf
A.\ref{causa}} and using the normalization of vacuum state.
\end{proof}

\begin{tre}
\label{acce}
The space of accessible states to a f-system $\sg \in \Sigma$ is given
by ${\cal H}_S \oplus {\cal H}_A \subset {\cal H}_E$ .
\end{tre}

\begin{proof}
See ref. \cite{Axiom2}.
\end{proof}

\mbox{}

\begin{remark}
From {\bf Thm. \ref{acce}} we deduce that non-symmetric states just
do not exist. This means that there is no arbitrary restriction to
symmetric or antisymmetric states.
\end{remark}

\subsubsection*{Group V: Symmetries and Group Structure}

\begin{ax}[Galilei Group]
\label{irred}
\begin{des}
\item {\bf [S]} $G \den$ Galilei Group.
\item {\bf [S]} $(\forall g)_G ( g=(b,{\bf a},{\bf v},R) \den $ the
general element of the Ga\-li\-lei group $)$.
\item {\bf [M]} The structure of Lie algebra $\tilde {\cal G}$ of the
extended group $\tilde G$, is generated by $\{ \hat H,\hat P_i,\hat
K_i,\hat J_i, \hat M \} \subset A $
\item {\bf [S]} $\hat H \den $ the time translations generator.
\item {\bf [S]} $(\forall \sg)_{\Sigma} ( eiv \hat H \den E \rep $ the
energy of $\sg)$.
\item {\bf [S]} $\hat P_i \den $ the spatial translations generator .
\item {\bf [S]} $(\forall \sg)_{\Sigma} ( eiv \hat P_i \den p_i \rep $
the i-th component of linear momentum of $\sg) $.
\item {\bf [S]} $\hat K_i \den $ the generator of the pure
transformations of Ga\-li\-lei.
\item {\bf [S]} $\hat J_i \den $ the generator of spatial rotations.
\item {\bf [S]} $(\forall \sg)_{\Sigma} ( eiv \hat J_i \den j_i \rep $
the i-th component of angular momentum of $\sg ) $.
\item {\bf [M]} $\hat M$ has a discrete spectrum of real eigenvalues
and its called {\em mass operator}.
\item {\bf [S]} $(\forall \sg)_{\Sigma} (eiv \hat M \den m \rep $ mass
of $\sg )$.
\item {\bf [P]} The vacuum state $|0 \rangle$ is the zero-mass state
that is invariant under Galilean transformations (up to a possible
phase factor accounting for a constant energy).
\end{des}
\end{ax}

\begin{ax}[Field Transformations]
\label{ftran}
\begin{des}
\item {\bf [M]} Let $\{ \hat \psi_\lambda({\bf x},t) ;
\lambda=-s,\ldots{},s \}$ be a set of field operators. They transform
locally under Galilean transformations $g=(b,{\bf a},{\bf v},R)$ as:
\begin{eqnarray*}
\hat U(g)\hat \psi_\lambda({\bf x},t) \hat U^{-1}(g) &=&
\exp[\frac{i}{\hbar} m \gamma(g;{\bf x},t)]\\
    && \sum_{\lambda'} D_{\lambda
\lambda'}^{(s)}(R^{-1})\hat \psi_{\lambda'}({\bf x}',t')
\end{eqnarray*}
where $D_{\lambda \lambda'}^{(s)}$ is the $(2s+1)$-dimensional unitary
matrix representation of the rotation group.
$$
\gamma(g;{\bf x},t)={\frac{1}{2}} {\bf v}^2 t + {\bf v} . R{\bf x},
$$
and
$$
{\bf x}' = R{\bf x} + {\bf v}t + {\bf a},
$$
$$
t'= t + b.
$$
\end{des}
\end{ax}

\begin{remarks}
\begin{enumerate}
\item The {\bf A.\ref{irred}} states the set of operators which are
the generators of the Lie algebra $\cal G$ of the Galilei group. The
elements of this set are semantically interpreted as specific
properties of the system.
\item The core of the GQFT is stated in {\bf A. \ref{ftran}}. This
axiom characterizes the transformation properties of the systems,
i.e. its Galilean symmetry properties. Note, moreover, that the
equation given in {\bf A. \ref{ftran}} may be seen like a {\em finite
interval field equation} in comparison with its differential form (see
{\bf Thm.  \ref{dife}}).
\end{enumerate}
\end{remarks}

\subsection{Theorems}
\begin{tre}[Mass conservation]
The Galilean invariance of a system requires  mass conservation.
\end{tre}

\begin{proof}
Since the operator $\hat M$ commutes with all the elements of the
extended Lie algebra $\tilde {\cal G}$, in particular with the
Hamiltonian $\hat H$.
\end{proof}

\begin{tre}[Bargmann superselection ``rule"]
\label{super}
${\cal H}$ decomposes into mutually orthogonal
vectors are eigenvectors of $\hat M$.
\end{tre}

\begin{proof}
From the extended algebra $\tilde {\cal G}$ and {\bf A. \ref{irred}}.
\end{proof}

\begin{tre} \label{dife}
The equation of motion satisfied by a give field operator is:
$$
i \hbar \frac{\partial}{\partial t}\hat \psi_\lambda= [\hat
\psi_\lambda ,\hat H]
$$
\end{tre}

\begin{proof}
Consequence of the invariance under time displacement.
\end{proof}

\begin{tre} \label{psidag}
\mbox{} The field operator $\hat \psi_\lambda^{\dag}({\bf x},t) $ transforms
locally under a Galilean transformation as:
$$
\hat U(g)\hat \psi_\lambda^{\dag}({\bf x},t) \hat U^{-1}(g)=
\exp[-\frac{i}{\hbar} m \gamma(g;{\bf x},t)] \sum_{\lambda'}
D_{\lambda' \lambda}^{(s)}(R)\hat \psi_{\lambda'}^{\dag}({\bf
x}',t').
$$
\end{tre}

\begin{proof}
From {\bf A. \ref{ftran}} and taking into account that $D_{\lambda
\lambda'}(R^{-1})=D_{\lambda \lambda'}^{\dag}(R)=(D_{\lambda'
\lambda}(R))^{*}$ since $D(R)$ is unitary.
\end{proof}

\begin{tre}
No Galilean field operator of non-zero mass can be hermitian.
\end{tre}

\begin{proof}
Compare {\bf Thm. \ref{psidag}} with the transformation rule of {\bf
A. \ref{ftran}}.
\end{proof}

\mbox{}

\begin{remarks}
\begin{enumerate}
\item The simplest of the representations of the Galilei group is
obtained for $D(R)=1$. This is the scalar representation of the
rotation group, which describes particles of zero spin.
\item {\bf Thm. \ref{super}} shows that the existence of a state which
is a superposition of two states with different masses, is in conflict
with Galilean invariance. This is a very strong constraint that does
not have a counterpart in relativistic theories since it restricts the
possible kind of process allowed by a Galilean theory.
\item Since the expanded Lie algebra is free of central charges and
has only ordinary representations, we are not forced to ``impose" any
superselection ``rule" \cite{We96}.
\item From the commutation relations of the Lie algebra ${\cal G}$
given in {\bf Def. \ref{liealg}}, we can see that the Hamiltonian
$\hat H$ does not appear on the right hand side. This means that in
order to obtain a covariant theory, the interaction Hamiltonian must
be invariant under a Galilean transformation, but the other generators
will remain the same as the free theory. This is to be compared with
the relativistic situation, where $\hat H$ appears in the Lie algebra
of Poincar\'e group, when the commutators of pure Lorentz and space
translation generators are taken. Thus, when the Hamiltonian is
modified by an interaction Hamiltonian, it requires a modification of
the pure Lorentz transformation, while no modification is needed in
the nonrelativistic case when the interaction Hamiltonian is Galilean
invariant. This explain why is so easy to construct nontrivial
Galilean theories \cite{JMLL67}.
\end{enumerate}
\end{remarks}

We shall now give some formal definitions motivated by the axioms
stated above:

\subsection{Definitions}

\begin{df}[Free particle state]
The state of a basic field characterized by an irreducible
representation of the Galilei group with null internal energy, denoted
by $[m,W=0,s]$, will be called {\em free particle state}.
\end{df}

\begin{df}[Total number operator]
\label{numoper}
The total number of particles operator is given by
$$
\hat N(t) = \sum_{\lambda} \int d^3 x {\hat \psi^{\dag}}_\lambda({\bf
x},t) \hat \psi_\lambda({\bf x},t) .
$$
\end{df}

\begin{df}[Free Hamiltonian]
The free Hamiltonian is given by
$$
\hat H_0=\sum_{\lambda}\int d^3 x {\hat \psi^{\dag}}_\lambda({\bf
x},t) (-\frac{\hbar^2}{2m}\nabla^2) \hat \psi_\lambda({\bf x},t) .
$$
\end{df}

\begin{remark}
Usually, the ``particle states" are called
``particles". We use the same convention in order to respect this
widely used designation.
\end{remark}

So far, the field operators describe only properties of free,
non-interacting fields. The multi-particle states encountered in free
field theories are eigenstates of the free Hamiltonian. This means
that they remain unchanged as long as the system is left
undisturbed. The following axiom relates the field operator associated
to interacting fields with the field operator associated to free
fields.

\subsubsection*{Group VI: Asymptotic Condition}
\begin{ax}
\begin{des}
\item {\bf [P]} Let $\hat H \in A$ such that $\hat H=\hat H_0+\hat V$,
with $\hat H \rightarrow \hat H_0$ as $t \rightarrow \pm \infty $,
where $\hat H_0 \den $ free Hamiltonian and $\hat V \den $ interaction
Hamiltonian.  Then, the field operators of $\hat \psi_\lambda({\bf
x},t) $ evolving with $\hat H$ behave as the free fields $\hat
\psi_\lambda^{in}({\bf x},t) $ or $\hat \psi_\lambda^{out}({\bf x},t)
$ for $t \rightarrow \pm \infty $, i.e.:
$$
\hat \psi_\lambda^{in}({\bf x},t) = \lim_{t \rightarrow -\infty} \hat
\psi_\lambda({\bf x},t) ,
$$
$$
\hat \psi_\lambda^{out}({\bf x},t) = \lim_{t \rightarrow +\infty} \hat
\psi_\lambda({\bf x},t) .
$$
\end{des}
\end{ax}

\begin{remarks}
\begin{enumerate}
\item This axiom is not valid for Coulombian potentials. To include
this kind of interaction in the theory we need to redefine the
operators $\hat H_0$ and $\hat V$ in order to satisfy the asymptotic
condition.
\item Redmond and Uretsky \cite{RU62}, showed that in the case of the so
called ``second quantization" the in-out operators cannot be defined
for Coulomb potentials or potential which decrease more slowly that
the Coulomb ones for large separations.
\end{enumerate}
\end{remarks}

The $S$-matrix can be defined in the same manner as in the
relativistic theory, from the following conventions:
\subsection{Definitions}
\begin{df}
\label{omeg}
Let $\hat H_0$ and $\hat H =\hat H_0 + \hat V$ be the free and full
Hamiltonians. We define the M{\"o}ller operator:
$$
\hat \Omega(t)\defi e^{+\frac{i}{\hbar}\hat H t}
e^{-\frac{i}{\hbar}\hat H_0 t}.
$$
\end{df}

\begin{df} [Evolution operator]
$$
\hat U(t,t_0) \defi \hat \Omega(t)^{\dag} \hat
\Omega(t_0)=e^{\frac{i}{\hbar}\hat H_0 t} e^{-\frac{i}{\hbar}\hat H (t
- t_0)} e^{-\frac{i}{\hbar}\hat H_0 t_0}.
$$
\end{df}

\begin{df}[Scattering operator]
\label{scaop}
$$
\hat S \defi \lim^{t \rightarrow +\infty}_{t_0 \rightarrow -\infty}
\hat U(t,t_0) .
$$
\end{df}

\begin{df}[Asymptotic states]
\label{asicon}
Let $| \alpha \rangle $ be the free-particle states and $\hat
\Omega(t)$ the M{\"o}ller operator.
\begin{description}
\item a) The {\em in-states} are defined as:
$
{| \alpha \rangle}_{in} \defi \lim_{t \rightarrow -\infty} \hat
\Omega(t) | \alpha \rangle.
$
\item b) The {\em out-states} are defined as:
$
{| \alpha \rangle}_{out} \defi \lim_{t \rightarrow +\infty} \hat
\Omega(t) | \alpha \rangle. 
$
\end{description}
\end{df}

\begin{df} [S-matrix]
\label{smatr}
Let us consider a process in which an initial configuration of
particles ${| \alpha \rangle}_{in} $ ends up as a final configuration
$ {|\beta \rangle}_{out} $. The elements of the S-matrix are given by
$$
S_{\beta \alpha} \defi _{out}{\langle \beta |}{\alpha
\rangle}_{in},
$$
and they represent the probability amplitude for the transition
$|\alpha \rangle \rightarrow |\beta \rangle $.
\end{df}

\begin{remark}
We shall agree to call {\em process} an evolution of field depending
on time. However, a formal definition of a process can be found in
\cite{BungeT3}. 
\end{remark}

\subsection{Theorems}
\begin{tre}[Normalization]
The in-states and out-states are normalized like the free particle
states; i.e.:
$$
_{in}{ \langle \alpha }| {\beta \rangle}_{in}=_{out}{ \langle \alpha
}| {\beta \rangle}_{out}= \langle \alpha |\beta \rangle = \delta(\beta
- \alpha).
$$
\end{tre}

\begin{proof}
From {\bf Def. \ref{asicon}} and taking into account that the states
are time independent.
\end{proof}

\begin{tre}
We can express the $\hat S$ operator between free particle states
$\langle \beta | $ and $|\alpha \rangle$ as:
$$
S_{\beta \alpha}= \langle \beta | \hat S \alpha \rangle.
$$
\end{tre}

\begin{proof}
From {\bf Def. \ref{scaop}} and {\bf Def. \ref{smatr}}.
\end{proof}

\section{GENERAL STRUCTURE OF GQFT}

In order to study the general structure of GQFT we shall now deduce
some of the specific theorems entailed by the axiomatic system. The
following conclusions were partially derived in ref.\cite{JMLL67,
DK62}.  The first two theorems restrain the possible kind of processes
allowed by Galilean invariance.

\subsection{Theorems}
\begin{tre}
\label{pairw}
Let $\{ {\hat \psi_{\lambda'_i}^{\dag}({\bf x}'_i,t)} , i=1\ldots{}N \}$
and $\{ {\hat \psi_{\lambda_j}({\bf x}_j,t)} , j=1\ldots{}M \}$ be the sets
of field operator associated to basic fields with the same mass
$m$. In order that the operator $\hat {\overline O}= \hat U(g) \hat O
\hat U^{-1}(g)$ given in {\bf A. \ref{oper}}, be invariant under
Galilean transformations, three conditions must be fulfilled:
\begin{description}
\item a) There are as many ${\hat \psi}^{\dag}$ as  ${\hat \psi}$
field operators, (i.e. $N=M$).
\item b) The ${\hat \psi}^{\dag}$ and ${\hat \psi}$ field operators
act pairwise at the same point.
\item c) The coefficients $C_{NM}$ are Galilean invariant, that is:
$$
C_{\overline \lambda'_1\ldots{}\overline \lambda'_N \overline
\lambda_1\ldots{}\overline \lambda_M}(\overline {\bf x}'_1,\ldots{},\overline
{\bf x}'_N,\overline {\bf x}_1,\ldots{},\overline {\bf x}_M) =
$$
$$
\sum_{\lambda'_1\ldots{}\lambda'_N} \sum_{\lambda_1\ldots{}\lambda_M}
D_{\overline \lambda'_1 \lambda'_1}(R)\ldots{}D_{\overline \lambda'_N
\lambda'_N}(R) D_{\lambda_1 \overline
\lambda_1}(R^{-1})\ldots{}D_{\lambda_M \overline \lambda_M}(R^{-1})
$$
$$
 \times C_{\lambda'_1\ldots{}\lambda'_N
\lambda_1\ldots{}\lambda_M}({\bf x}'_1,\ldots{},{\bf x}'_N, {\bf
 x}_1,\ldots{},{\bf 
x}_M) .
$$
\end{description}
\end{tre}

\begin{proof}
See Appendix.
\end{proof}

\begin{tre}[Conservation of the number of particles]
\label{conpar}
\mbox{} The operator $\hat O$ given in {\bf Thm. \ref{pairw}} commutes
with the number operator $\hat N(t)$.
\end{tre}

\begin{proof}
From $a)$ and $b)$ of {\bf Thm. \ref{pairw}} and {\bf Def.\ref{numoper}}.
\end{proof}

\begin{tre}
\label{nocon}
Let $\{\hat \psi_{\lambda'_i}^{\dag}({\bf x},t) , i=1\ldots{}N \}$ and
$\{\hat \psi_{\lambda_j}({\bf x},t) , j=1\ldots{}M \}$ be the sets of
field operators associated to basic fields with {\em different} masses
denoted by $m'_i$ and $m_j$ respectively and taken at the same point
of space-time. The operator $\hat {\overline O}$ will be invariant
under Galilean transformations if
$$
\sum_i^N m'_i = \sum_j^M m_j.
$$
\end{tre}

\begin{proof}
See Appendix.
\end{proof}

\begin{tre}[Non-conservation of the number of particles]
\label{nocopar}
\mbox{} The operator $\hat {\overline O}$ given in {\bf
Thm. \ref{nocon}} does not commute with the number operator.
\end{tre}

\begin{proof}
From the fact that in general $N \ne M$.
\end{proof}

\mbox{}

\begin{remarks}
\begin{enumerate}
\item As an example of {\bf Thm. \ref{pairw}} we have the ``two-body"
interaction Hamiltonian $\hat V$ which describes the mutual action
between two ``particles",
$$
\hat V=\frac{1}{2}\int d^3 x d^3 x' {\hat \psi^{\dag}}({\bf x},t)
{\hat \psi^{\dag}}({\bf x}',t) \hat V(|{\bf x}-{\bf x}'|) \hat
\psi({\bf x}',t) \hat \psi({\bf x},t)
$$
In this way,  second quantization appears as a particular case of
another theory when certain subsidiary assumptions are adjoined.
\item From {\bf Thm. \ref{pairw}} and {\bf Thm. \ref{conpar}} we can see
that in a theory with only one kind of particles of fixed mass, mass
conservation necessarily implies conservation of the number of
particles and forbids production processes \cite{DK62}.
\item As we can see from {\bf Thm. \ref{nocon}} and {\bf
Thm. \ref{nocopar}} it is possible to have Galilean invariant theories
that describe production processes in which the number of particles is
not conserved but the total mass is conserved. In this case we have to
include field operators associated to several kinds of basic fields of
different masses and acting at the same point \cite{JMLL67}.
\item {\bf Thm. \ref{nocon}} provides a rule for building a theory that
describe production processes. It means that the field operators must
be chosen in agreement with the mass conservation as expressed in {\bf
Thm. \ref{nocopar}}. This is a strong constraint and it does not exist
in relativistic field theory.
\end{enumerate}
\end{remarks}

\subsection{Definitions}

We inquire now whether the spin-statistics relation and crossing
symmetry are required by GQFT. In order to answer this question, we
need to introduce the concept of an ``antiparticle". This concept is
also motivated by the possibility of describing ``pair production"
processes, as can be seen from {\bf Thm. \ref{nocon}}. Thus, this
model allows us to interpret particles with negative mass eigenvalues
as an antiparticle, i.e.,
\begin{df}[Free Antiparticle state]
The state of a basic field characterized by an irreducible
representation of the Galilei group with null internal energy, denoted
by $[-m, -W=0,s^*]$, where $s^*$ denote the complex conjugate
representation of $D(R)$ will be called a {\em free antiparticle
state}.
\end{df}

According to the Peter-Weyl theorem, the functions $D_{\lambda
\lambda'}^{(s)}(R)$ form a complete basis in the space of square
integrable functions (or operator valued functions). For the
non-compact part of the Galilei group we shall use the fact that we
are dealing with a rigged Hilbert space ${\cal H}_E$, where the
eigenfunctions of infinite norm form a complete set. In this way, we
can take the next definition (usually referred as the {\em plane wave
expansion} of the field operator) as ``analogous" to the Peter-Weyl
theorem for non-compact groups. Thus, we can define the following field
operators:
\begin{df}[Annihilation field operator]
\label{destf}
The free field operator associated with a basic field $\sg_i$ given
by:
\begin{displaymath}
\hat \psi^-_\lambda({\bf x},t) \defi (2\pi)^{-3/2}\sum_{\lambda'} \int
d\mu({\bf p},E) e^{\frac{i}{\hbar} (Et - {\bf {\bf p.x}})} D_{\lambda
\lambda'}^{(s)}(R^{-1}) \hat a({\bf p},E,\lambda').
\end{displaymath}
will be called the ``annihilation" field operator of free particles.
\end{df}

We can define a new field operator by taking the hermitian conjugate
of the field defined above,
\begin{df}[Creation field operator]
\label{creaf}
The free field operator associated a basic field $\sg_i$ given by:
$
\hat \psi^+_\lambda({\bf x},t) \den {\hat \psi^{-\dag}_\lambda}({\bf
x},t) \defi$
$$ (2\pi)^{-3/2} \sum_{\lambda'} \int d\mu({\bf p},E)
e^{-\frac{i}{\hbar}(Et - {\bf p.x})} D_{\lambda' \lambda}^{(s)}(R)
\hat a^{\dag}({\bf p},E,\lambda') $$
will be called the ``creation" field operator of free particles.
\end{df}

\begin{remarks}
\begin{enumerate}
\item As it is well known, the mass-shell condition $p^2+m^2=0$ in the
relativistic case provides negative and positive solutions. This is
related to the existence of an antiparticle state for every ordinary
particle state, where we assume that one particle is the antiparticle
of another if their masses and spin are equal and their charges are
opposite. In the nonrelativistic case the eigenvalues of $\hat M$ can
be positive or negative, and we can interpret to a particle with
positive mass-eigenvalue and the antiparticle with the opposite
mass-eigenvalue. So the mass behaves in nonrelativistic case as a kind
of charge. This is due to the fact that there exists a ``Bargmann's
superselection rule".
\item We won't assume that every particle has an a antiparticle since
this must be proved from our axioms. In the case of particles that are
their own antiparticles we take $\hat b(q)=\hat a(q)$.
\item Since quantum field theory deals with transmutations of systems,
the operators given in {\bf Def. \ref{destf}} and {\bf Def. \ref{creaf}}
conceptualize the change brought about by the decrement or increment
in the number of particles of the system.
\end{enumerate}
\end{remarks}

\subsection{Theorems}

\begin{tre} \label{comm}
The operator
coefficients $\hat a(k)\den \hat a({\bf p},E,\lambda)$, $\hat
a^{\dag}(k) \den \hat a^{\dag}({\bf p},E,\lambda)$ satisfy
$$
[\hat a(k'),\hat a^{\dag}(k)]_{\mp}= \delta (k'-k) ,
$$
$$
[\hat a(k'),\hat a(k)]_{\mp}=[\hat a^{\dag}(k'),\hat
a^{\dag}(k)]_{\mp}=0.
$$
\end{tre}

\begin{proof}
From  {\bf A. \ref{causa}}, {\bf Def. \ref{creaf}} and {\bf
Def. \ref{destf}}.
\end{proof}

\begin{tre}\mbox{}
\label{crea}
The operators $\hat a^{\dag}(k)$ and $\hat a(k)$ that satisfy the
above commutation relations act on a general state $|k_1,
k_2,\ldots{}, k_N \rangle$ as:
\begin{description}
\item a) $\hat a^{\dag}(k) |k_1, k_2,\ldots{}, k_N \rangle =|k, k_1,
k_2,\ldots{} ,k_N \rangle$.\
\item b) $\hat a(k) |k_1, k_2,\ldots{}, k_N \rangle = \sum_{r=1}^N
(\pm)^{r+1} \delta(k-k_r) |k_1, \ldots{} ,k_{r-1},k_{r+1}, \ldots{} ,k_N
\rangle$\\
 with $+1$ or $-1$ for\\
$|k_1, k_2,\ldots{},k_N \rangle \in {\cal H}_S$ and
${\cal H}_A$ respectively.
\end{description}
\end{tre}

\begin{proof}
From {\bf Thm. \ref{gral}}, {\bf Def. \ref{destf}} and {\bf
Def. \ref{creaf}}.
\end{proof}

\begin{tre} \label{autoad}
The operators $\hat a$ and $\hat a^{\dag}$ are mutually adjoint:
$(\hat a^{\dag}(k))^{\dag} = \hat a(k)$.
\end{tre}

\begin{proof}
Using {\bf Def. \ref{destf}} and {\bf Def. \ref{creaf}}.
\end{proof}

\begin{tre}
\label{trans}
\mbox{} The operator $\hat a^{\dag}({\bf p},E,\lambda)$ transforms under an
arbitrary Galilei transformation according to:
\begin{displaymath}
\hat U(g) \hat a^{\dag}({\bf p},E,\lambda) \hat
U^{-1}(g)=e^{-\frac{i}{\hbar}(E' b -{\bf p'.a})} \sum_{\lambda'}
D_{\lambda' \lambda}^{(s)}(R) \hat a^{\dag}({\bf p}',E',\lambda')
\end{displaymath}
with ${\bf p}'$, $E'$ and $D^{(s)}(R)$ as given in {\bf Thm. \ref{repre}}.
\end{tre}

\begin{proof}
From {\bf Thm. \ref{crea}} and {\bf
Thm. \ref{repre}}. (Alternatively: from {\bf A. \ref{ftran}} and {\bf
Def. \ref{creaf}}).
\end{proof}

\begin{tre}
\label{transa}
The operator $\hat a({\bf p},E,\lambda)$ transforms under a Galilei
transformation as:
$$
\hat U(g) \hat a({\bf p},E,\lambda) \hat U^{-1}(g)=e^{\frac{i}{\hbar}
(E' b -{\bf p'.a})} \sum_{\lambda'} D_{\lambda \lambda'}^{(s)}(R^{-1})
\hat a({\bf p}',E',\lambda') .
$$
\end{tre}

\begin{proof}
Using {\bf Thm. \ref{transa}} and {\bf Thm. \ref{autoad}}.
\end{proof}

\mbox{}

In order to put the transformation of $\hat \psi_{\lambda}^{\dag}({\bf
x},t) $ in a similar form as $\hat \psi_{\lambda}({\bf x},t) $, it
will be convenient to consider the following theorem (we follow the
same idea given by \cite{Wien64} for the relativistic case),
\begin{tre}
\label{alpha}
 The operators $ \hat \alpha ({\bf p},E,\lambda) \defi \sum_{\lambda'}
D_{\lambda 
\lambda'} (R^{-1}) \hat a({\bf p},E,\lambda')  $ and\\
 $ \hat \beta^{\dag} ({\bf
p},E,\lambda) \defi \sum_{\lambda'} {\{D(R^{-1})C^{-1}\}}_{\lambda \lambda'}
\hat b^{\dag}({\bf p},E,\lambda')  $ transform as:
$$
\hat U(g) \hat \alpha({\bf p},E,\lambda) \hat
U^{-1}(g)=e^{\frac{i}{\hbar}(E' b -{\bf p'.a})} \sum_{\lambda'}
D_{\lambda \lambda'}^{(s)}(R^{-1}) \hat \alpha ({\bf p}',E',\lambda')
,
$$
$$
\hat U(g) \hat \beta^{\dag}({\bf p},E,\lambda) \hat
U^{-1}(g)=e^{-\frac{i}{\hbar}(E' b -{\bf p'.a})} \sum_{\lambda'}
D_{\lambda \lambda'}^{(s)}(R^{-1}) \hat \beta^{\dag}({\bf
p}',E',\lambda').
$$
\end{tre}

\begin{proof}
The transformation of $\hat \alpha({\bf p},E,\lambda)$ follows from
{\bf Thm.\ref{transa}}. To obtain $\hat \beta^{\dag}({\bf
p},E,\lambda)$, use the {\bf Thm. \ref{trans}} and the following
properties of the unitary representation of $D(R)$:
$$
(D^{(s)}(R))^*=C D^{(s)}(R) C^{-1}
$$
so, we can write:
$$
D_{\lambda'\lambda}^{(s)}(R)=(D_{\lambda\lambda'}^{(s)}(R^{-1}))^*=(C
D^{(s)}(R^{-1}) C^{-1})_{\lambda\lambda'}
$$
where $C$ is a $(2s+1) \times (2s+1)$ matrix with
$$
C^*C=(-1)^{2s} \;C^{\dag}C=1
$$
\end{proof}

\begin{tre}
\label{transc}
The creation field operator of antiparticles given by:
$$
\hat \psi_{\lambda}^{-c \dag}({\bf x},t) \defi (2\pi)^{-3/2} \int
d\mu({\bf p},E) e^{-\frac{i}{\hbar}(Et - {\bf p.x})} \hat
\beta^{\dag}({\bf p},E,\lambda)
$$
transforms as required by {\bf A. \ref{ftran}}, i.e.:
\begin{eqnarray*}
\hat U(g) \hat \psi_\lambda^{-c \dag}({\bf x},t) \hat U^{-1}(g) &=& \exp
[-\frac{i}{\hbar}(-m) \gamma(g;{\bf x},t)] \sum_{\lambda'} D_{\lambda
\lambda'}^{(s)}(R^{-1}) \hat \psi_{\lambda'}^{-c \dag}({\bf x}',t').
\end{eqnarray*}
\end{tre}

\begin{proof}
Using {\bf Thm. \ref{alpha}} (See Appendix for details).
\end{proof}

\mbox{}

In order to obtain Hermitian operators and field operators that
satisfy the transformation rule of {\bf A.\ref{ftran}} we construct a
field operator by taking linear combinations of $\hat
\psi_\lambda^-({\bf x},t) $ and $\hat \psi_\lambda^{-c \dag}({\bf
x},t) $. That is, we need the following definition:

\begin{df}\label{gener}
The local field operators constructed as linear combinations of
particles annihilation field operator and antiparticles creation field
operator are given by:
$$
\hat \psi_\lambda({\bf x},t) = \xi \hat \psi_\lambda^-({\bf x},t) +
\eta \hat \psi_\lambda^{-c \dag}({\bf x},t)
$$
$$
=(2\pi)^{-3/2} \int d\mu({\bf p},E) \{ \xi e^{\frac{i}{\hbar}(Et-{\bf
p.x})} \hat \alpha({\bf p},E,\lambda) + \eta
e^{-\frac{i}{\hbar}(Et-{\bf p.x})} \hat \beta^{\dag}({\bf
p},E,\lambda) \}.
$$
\end{df}

\begin{tre}
The field operator $\hat \psi_\lambda({\bf x},t)$ as given in {\bf Def.
\ref{gener}} has the property of Galilei transformation given in {\bf A.
\ref{ftran}}.
\end{tre}

\begin{proof}
From {\bf Def. \ref{gener}} and using {\bf Thm. \ref{alpha}} and {\bf
Thm. \ref{transc}}.
\end{proof}

\begin{tre}
\label{comgen}
The field operator $\hat \psi_\lambda({\bf x},t)$ as given in {\bf
Def. \ref{gener}} satisfies the
commutation or anticommutation rule:
$$
[\hat \psi_\lambda({\bf x},t) ,\hat \psi_{\lambda'}^{\dag}({\bf
y},t)]_{\mp}= (|\xi|^2 \mp |\eta|^2 ) \delta_{\lambda \lambda'}
\delta^3 ({\bf x}-{\bf y}).
$$
\end{tre}

\begin{proof}
Using {\bf A. \ref{causa}}.
\end{proof}

\mbox{}

Thus, this theorem leads immediately to the two most important
consequences of the Galilean theory,
\begin{tre}[Crossing and Statistics]
\label{stat}
\begin{description}
\item[Statistics:] No statistics can be deduced from a Galilean
theory field.
\item[Crossing:] A full crossing symmetry with $|\xi|=|\eta|$ is not
required.
\end{description}
\end{tre}

\begin{proof}
The commutation rule of {\bf Thm. \ref{comgen}} satisfies the local
commutability condition given in {\bf A. \ref{causa}} for any value of
$\xi$ and $\eta$.
\end{proof}

\mbox{}

\begin{remarks}
\begin{enumerate}
\item It is easy to see that the field operator of {\bf
Def. \ref{gener}} cannot be built as a linear combination of $\hat a$
and $\hat a^{\dag}$ because it does not transform as required by {\bf
A. \ref{ftran}}, since $\hat a$ and $\hat a^{\dag}$ annihilate and
create the same particle with mass $m$.
\item The case of a {\em particle} without {\em antiparticle} is
satisfied by $\eta=0$.
\item {\bf Thm. \ref{stat}} does not select a sign for the commutator
(i.e.$\mp$) in order to describe a particle with integer or
half-integer spin. In other words, from the axiom of local
commutability ({\bf A. \ref{causa}}) we cannot decide whether a
particle with integer (or half-integer) spin must be a boson or a
fermion. Thus, Galilean field theories do not imply any relation
between spin and statistics.
\item There is no symmetry in GQFT between particle and
antiparticle. Indeed, any value of $\xi$ and $\eta$ are admissible,
even the non existence of antiparticles. So, CPT theorem has no
analogous in GQFT.
\item From {\bf Thm. \ref{stat}} we can appreciate the weakening of the
commutation relations of Galilean quantum field theory since in
relativistic quantum field theory, crossing symmetry and the relation
between spin and statistic arises from the ``caus\-al\-ity" requirement,
as has been shown by Weinberg \cite{Wien64}.
\end{enumerate}
\end{remarks}

\section{CONCLUSIONS}

We have presented a formal axiomatization of the Galilean quantum
field theories. These theories are generated by ten primitive
concepts, among them the reference class, i.e., fields. The
constraints are obtained from the choice of a symmetry group plus
appropriate commutation rules. This means that we consider the fields
as things with properties represented by operators that satisfy
certain symmetry transformations. Fields are unobservable but they should
not be regarded as auxiliary
devices with no physical meaning. Indeed, due to its success, the
program of explaining the behavior of matter in terms of fields
should be considered as an approximately true model of reality. On the
other hand, free particles
are described as states of the basic fields characterized by an
irreducible representation of the Galilei group with null internal
energy. This does not imply that particles are consequences of
symmetries since symmetries are symmetries of properties of things. In
other words, no things, no symmetries.

As a comparison,  second quantization is a generalization of
quantum mechanics to the case of $n$ ``particles" and differs of a
quantum field theory in the primitive concepts. In this theory, the
primitive concept is the quantum field as an ``extended object" whose
elementary excitations describe a ``quanton". This last real object is
the fundamental primitive of quantum mechanics of systems (see
reference \cite{Axiom1,Axiom2}).

The existence of non-trivial Galilean quantum field theories is
asserted by the ``Gali-Lee'' model and other examples studied by
L\'evy-Leblond \cite{JMLL67}. Moreover, non-relativistic second
quantization itself can be interpreted as a Galilean quantum field
theory, whose elementary excitations are ``quanton-like'' objects.

The general structure of GQFT has a number of interesting physical
consequences  that can be
resumed as follows:
\begin{enumerate}
\item From the property of field transformations, the spin of the
particles can be described in a Galilean frame.
\item Mass conservation is obtained from the assumption of Galilei
group as symmetry group of the theory. Moreover, Bargmann's
superselection rule is obtained as a theorem that prevents the
superposition of states of different masses.
\item Galilean invariant operators describing processes with number of
particles conservation of the same mass must be constructed starting
from creation and annihilation operators acting pairwise at the same
point.
\item In spite of the fact that the Bargmann's superselection rule
imposes strong constraints, production process with non-conservation
of number of particles are admitted in a Galilean theory. To this end,
the field operators must be associated to particles of different mass.
\item In order to describe pair production, antiparticles with
negative mass-eigenvalue can be defined. Thus, the counterpart to the
par\-ti\-cle-an\-ti\-par\-ti\-cle relation is not lost in a
nonrelativistic frame.
\item However, no crossing symmetry between particle and antiparticle
as in a relativistic theory can be deduced.
\item Neither can the spin statistics relation be derived from local
commutability. So, the CPT theorem has not counterpart in a Galilean
theory.
\end{enumerate}

Let us comment very briefly some philosophical consequences of the
pres\-ent axiomatization. Using the semantic tools defined in reference
\cite{BungeT1} one can identify from our axiom system the physical
reference class of the theory, i.e., the set of physical objects it
describes. The reference class is $\mathcal{R} = \langle \Sigma \times
\bar{\Sigma} \rangle$, i.e. quantum field systems and their
environments.  There are no ``observers'' or ``experiments'' in the
interpretation of either the axioms or the theorems of the theory
although the latter (but not the former) can be analyzed within the
theory using suitable additional hypothesis. This results shows the
realistic character of the theory. Note that the reference class can
be identified only when the axioms, background and primitive basis of
the theory have been stated in a formal way.

To summarize, the benefit of constructing this axiomatization becomes
clear since it provides a deeper understanding of Galilean Quantum
Field Theories and shows that they have a very rich structure. We
think that the consequences of this structure is a valuable tool to
``measure" the stronger results coming from relativistic
theory. However, this task cannot be make completely until its
realistic axiomatization that will be presented in a future paper.

\section*{ACKNOWLEDGMENTS}
The authors would like to thank S. Perez Bergliaffa for helpful
comments, M. A. Bunge and L. de la Pe\~na for a critical reading of
the manuscript and important advices. HV acknowledges support of
Project \texttt{G130} of University of La Plata, Argentina and Project
\texttt{No.  42026-F} of CONACYT, Mexico and . GP acknowledges support
from FOMEC scholarship program.

\section*{Appendix}

\subsection*{Proof of \bf Thm. \protect \ref{pairw}}
The transformation  of the operator can be written as:
$$
\hat {\overline O}= \hat U(g) \hat O \hat U^{-1}(g) =\sum_{N,M
=0}^{\infty}\sum_{ \lambda'_1\ldots{}\lambda'_N}
\sum_{\lambda_1\ldots{}\lambda_M} \int {\cal D} {\bf x}'{\cal D} {\bf x}
\hat U(g) {\hat \Psi}_{\lambda'}^{\dag}( {\bf x}', t) U^{-1}(g) \times
$$
$$
\hat U(g) {\hat \Psi}_{\lambda}({\bf x}, t) U^{-1}(g) {\cal C}_{NM}
$$
where
$$
\hat \Psi_{\overline \lambda'}^{\dag}(\overline {\bf x}',\overline
t)=\hat U {\hat \Psi}_{\lambda'}^{\dag}({\bf x}',t) \hat U^{-1}=
\exp\{-\frac{i}{\hbar} m [\frac{1}{2} {\bf v}^2 N t + {\bf v}. R ({\bf
x}'_1+\ldots{}+{\bf x}'_M)] \} \times
$$
$$
\sum_{\overline \lambda'_1\ldots{}\overline \lambda'_N} D_{\overline
\lambda'_1 \lambda'_1}(R)\ldots{}D_{\overline \lambda'_N
\lambda'_N}(R){\hat \psi}_{\overline \lambda'_1}^{\dag}(\overline {\bf
x}'_1,\overline t)\ldots{}{\hat \psi}_{\overline
\lambda'_N}^{\dag}(\overline {\bf x}'_N,\overline t)
$$
$$
\hat \Psi_{\overline \lambda}(\overline {\bf x},\overline t)=\hat U
{\hat \Psi}_{\lambda}({\bf x},t) \hat U^{-1}= \exp\{\frac{i}{\hbar} m
[\frac{1}{2} {\bf v}^2 M t + {\bf v}. R ({\bf x}_1+\ldots{}+{\bf x}_M)] \}
\times
$$
$$
\sum_{\overline \lambda_1\ldots{}\overline \lambda_M} D_{\overline
\lambda_1 \lambda_1}(R^{-1})\ldots{}D_{\overline \lambda_M
\lambda_M}(R^{-1}){\hat \psi}_{\overline \lambda_1}^{\dag}(\overline
{\bf x}_1,\overline t)\ldots{}{\hat \psi}_{\overline
\lambda_M}^{\dag}(\overline {\bf x}_M,\overline t)
$$
Replacing the last two expressions in $\hat{\overline O}$ we obtain:
$$
\hat {\overline O}= \hat U(g) \hat O \hat U^{-1}(g)=\sum_{N,M
=0}^{\infty}\sum_{\overline \lambda'_1\ldots{}\overline \lambda'_N}
\sum_{\overline \lambda_1\ldots{}\overline \lambda_M} \int {\cal
D}\overline {\bf x}'{\cal D}\overline {\bf x} {\hat \Psi}_{\overline
\lambda'}^{\dag}(\overline {\bf x}',\overline t){\hat \Psi}_{\overline
\lambda}(\overline {\bf x},\overline t) \overline {\cal C}_{NM}
$$
 where (a) and (b) follow from the requirement that the phase factors
 cancels. (c) follows immediately.

\subsection*{Proof of \bf Thm. \protect \ref{nocon}}
The transformed operator takes the form:
$$
\hat {\overline O}=\sum_{N,M =0}^{\infty}\sum_{\overline
\lambda'_1\ldots{}\overline \lambda'_N} \sum_{\overline
\lambda_1\ldots{}\overline \lambda_M} \int {\cal D}\overline {\bf x} {\hat
\Psi}_{\overline \lambda'}^{\dag}(\overline {\bf x},\overline t){\hat
\Psi}_{\overline \lambda}(\overline{\bf x},\overline t) \overline
{\cal C}_{NM}
$$
with
$$
{\hat \Psi}_{\overline \lambda'}^{\dag}(\overline {\bf x},\overline t
) = \exp [ - \frac{i}{\hbar} (m'_1+\ldots{}+m'_N) ( \frac{1}{2} {\bf v}^2
t
+ {\bf v}.R {\bf x})] D_{\overline \lambda'_1 \lambda'_1}(R)\ldots{}
$$
$$
\ldots{}D_{\overline \lambda'_N \lambda'_N}(R){\hat \psi}_{\overline
\lambda'_1}^{\dag}(\overline {\bf x},\overline t)\ldots{}{\hat
\psi}_{\overline \lambda'_N}^{\dag}(\overline {\bf x},\overline t)
$$
$$
{\hat \Psi}_{\overline \lambda}(\overline {\bf x},\overline t ) = \exp
[ \frac{i}{\hbar} (m_1+\ldots{}+m_M) (\frac{1}{2} {\bf v}^2 t + {\bf v}. R
{\bf x}) ] D_{\overline \lambda_1 \lambda_1}(R^{-1})\ldots{}
$$
$$
\ldots{}D_{\overline \lambda_M \lambda_M}( R^{-1}){\hat \psi}_{\overline
\lambda_1}^{\dag}(\overline {\bf x},\overline t)\ldots{}{\hat
\psi}_{\overline \lambda_M}^{\dag}(\overline {\bf x},\overline t)
$$
The theorem follows from the above expressions with the requirement of
phase factors cancels.

\subsection*{Proof of \bf Thm. \protect \ref{transc}}
By steps:
\begin{description}
\item a) First, we prove the following expression:
$$
e^{-\frac{i}{\hbar}(E t - {\bf p}.{\bf x})} e^{-\frac{i}{\hbar}( E' b
- {\bf p}'.{\bf a})} = e^{-\frac{i}{\hbar}(E't' - {\bf p}'.{\bf x}')}
e^{-\frac{i}{\hbar}\tilde m \gamma}
$$
with $\gamma = \frac{1}{2}{\bf v}^2 t + {\bf v}. R{\bf x}$.  Using the
definitions of $E$ and ${\bf x}$, we can write the LHS as:
$$
\mbox{LHS}= e^{-\frac{i}{\hbar}[(E'-{\bf v}. R{\bf p}-\frac{1}{2}\tilde m {\bf
v}^2) t - R{\bf p}.({\bf x}'- {\bf v}t - {\bf a})]}
e^{-\frac{i}{\hbar}( E' b - {\bf p}'.{\bf a})}
$$
then inserting the expression of $t$ and arranging the terms, we
obtain:
$$
\mbox{LHS}=e^{-\frac{i}{\hbar}[E'(t'-b) - {\bf v}.R {\bf p}(t'-b) -
\frac{1}{2}\tilde m {\bf v}^2 t - R {\bf p}. {\bf x}' + R {\bf p}.{\bf
v}(t'-b) + R {\bf p}.{\bf a} + E' b - {\bf p}'.{\bf a}]}
$$
doing the same for ${\bf p}$ and ${\bf x}'$ and canceling we have,
\begin{eqnarray*}
\mbox{LHS} &=& e^{-\frac{i}{\hbar}[E't'- \frac{1}{2}\tilde m {\bf v}^2 t - ({\bf
p}'- \tilde m {\bf v}). {\bf x}'+ ({\bf p}'- \tilde m {\bf v}). {\bf
a} - {\bf p}'.{\bf a}]} \cr
&=& e^{-\frac{i}{\hbar}[E' t' -{\bf p}'.{\bf x}'- \frac{1}{2}\tilde m
{\bf v}^2 t + \tilde m {\bf v}.(R{\bf x}+ {\bf v} t + {\bf a}) -
\tilde m {\bf v}.{\bf a}]}
\end{eqnarray*}
and using the expression of $\gamma$, we arrive to:
$$
\mbox{LHS}=e^{-\frac{i}{\hbar}[E't'-{\bf p}'.{\bf x}'+ \frac{1}{2}\tilde m {\bf
v}^2 t + \tilde m {\bf v}.R{\bf x}]}= e^{-\frac{i}{\hbar}(E't' - {\bf p}'.{\bf x}')}
e^{-\frac{i}{\hbar}\tilde m \gamma}
$$
\item b) Now, we prove the transformation rule:
$$
\hat U(g) \hat \psi_{\lambda}^{-c \dag}({\bf x},t) \hat U^{-1}(g) =
(2\pi)^{-3/2} \int d \mu e^{-\frac{i}{\hbar}(E t - {\bf p}.{\bf x})}
\hat U(g) \hat \beta^{\dag}({\bf p},E,\lambda) \hat U^{-1}(g)
$$
from {\bf Thm. \ref{alpha}},
\begin{eqnarray*}
\hat U(g) \hat \psi_{\lambda}^{-c \dag}({\bf x},t)\hat U^{-1}(g) &=& (2\pi)^{-3/2}\int d\mu
e^{-\frac{i}{\hbar}(Et - {\bf p}.{\bf x})} e^{-\frac{i}{\hbar}(E' b - {\bf p}'.{\bf a})}
\times \cr
&& \sum_{\lambda'} D_{\lambda \lambda'}^{(s)}(R^{-1}) \hat \beta^{\dag}({\bf
p}',E',\lambda')
\end{eqnarray*}
using the expression obtained previously and arranging, we have:
\begin{eqnarray*}
\hat U(g) \hat \psi_{\lambda}^{-c \dag}({\bf x},t) \hat U^{-1}(g) &=& (2\pi)^{-3/2}
\int d\mu e^{-\frac{i}{\hbar}(E't' - {\bf p}'.{\bf x}')} e^{-\frac{i}{\hbar} \tilde m
\gamma } \times \cr
&& \sum_{\lambda'} D_{\lambda \lambda'}^{(s)}(R^{-1}) \hat \beta^{\dag}({\bf
p}',E',\lambda')
\end{eqnarray*}
$$
=e^{-\frac{i}{\hbar} \tilde m \gamma } \sum_{\lambda'} D_{\lambda
\lambda'}^{(s)}(R^{-1}) [(2\pi)^{-3/2} \int d\mu
e^{-\frac{i}{\hbar}(E't' - {\bf p}'.{\bf x}')} \hat \beta^{\dag}({\bf
p}',E',\lambda') ]
$$
using the fact that the antiparticle has mass $\tilde m = - m$ and the
definition of $\hat \psi_{\lambda'}^{-c \dag}({\bf x}',t')$ we obtain
finally,
$$
\hat U(g) \hat \psi_{\lambda}^{-c \dag}({\bf x},t) \hat U^{-1}(g)
=e^{-\frac{i}{\hbar}(-m) \gamma} \sum_{\lambda'} D_{\lambda
\lambda'}^{(s)}(R^{-1}) \hat \psi_{\lambda'}^{-c \dag}({\bf x}',t')
$$
\end{description}

\end{document}